\documentclass[fleqn,twoside]{article}
\usepackage[headings]{espcrc2}

\readRCS
$Id: espcrc2.tex,v 1.2 2004/02/24 11:22:11 spepping Exp $
\usepackage{graphicx}
\usepackage{epsfig}
\usepackage[figuresright]{rotating}

\newcommand{\AmS}{{\protect\the\textfont2
  A\kern-.1667em\lower.5ex\hbox{M}\kern-.125emS}}

\title{ 
{\normalsize \tt
IFJPAN-V-2005-10}\\
\vspace*{1.cm}
\bf Multi-Particle Processes in QCD without Feynman Diagrams}

\author{ Costas G. Papadopoulos\address{Institute of Nuclear Physics, 
        NCSR Demokritos, 15-310 Athens, Greece} and
        Ma\l gorzata  Worek\addressmark\address{Institute of Nuclear 
        Physics PAS, Radzikowskiego 152, 31-3420, Cracow, 
        Poland}\thanks{Presented at the X International Workshop on Advanced 
        Computing and Analysis Techniques in Physics Research, ACAT 2005, 
        DESY-Zeuthen, Germany, 22-27 May 2005.}}

\runtitle{Calculation of Multi-Particle Processes in QCD}
\runauthor{C.G. Papadopoulos and M. Worek}

\begin{document}

\begin{abstract}
A way to efficiently compute helicity amplitudes for 
arbitrary tree-level scattering processes in QCD is presented. 
The scattering amplitude is evaluated recursively 
through a set of Dyson-Schwinger equations. The computational cost of this 
algorithm grows asymptotically as $3^n$, where $n$ is the number of particles 
involved in the process, compared to $n!$ in the 
traditional Feynman graphs approach. Unitary gauge is used and mass effects are available as well. Additionally, 
the color and helicity structures are appropriately transformed so the usual 
summation is replaced by the Monte Carlo techniques.
\vspace{1pc}
\end{abstract}

\maketitle

\section{INTRODUCTION}

QCD processes with many external legs are of much interest, both for testing 
QCD in different settings and as backgrounds for
new physics processes at the Fermilab TeVatron and at the CERN LHC. 
However, the estimation of multi-jet production
cross sections as well as their characteristic distributions is a difficult
task. Perturbation theory based on Feynman graphs runs into computational
problems, since the number of graphs contributing to the amplitude grows
like $n!$. The counting of the graphs itself becomes a problem let alone
their evaluation and the computation of the color and helicity structures
which is an additional source of computational inefficiencies. Over the last
few years, new  recursive algorithms based on Dyson-Schwinger 
equations \cite{Kanaki:2000ey,Kanaki:2000ms,Draggiotis:2002hm} 
or on field equations 
\cite{Berends:1987cv,Caravaglios:1995cd,Draggiotis:1998gr,Caravaglios:1998yr}
have been developed in order to overcome the computational 
obstacles. Very recently also on shell recursive equations have been
proposed \cite{Britto:2004ap,Britto:2005fq}.
The amplitude calculated using Dyson-Schwinger recursive equations, 
which avoid Feynman diagrams, results in a computational 
cost growing asymptotically as $3^n$. Here, the off-shell subamplitudes
 are introduced which
are combinations of parts of Feynman graphs.
For those subamplitudes a recursion relation has been obtained which enables to
express an n-particle amplitude in terms of all subamplitudes, 
starting from $1-$, $2-$, ... up to $(n-1)$ particles. 
The color and helicity structures corresponding to each
subamplitude have a simpler form. Moreover, they can be
appropriately transformed so the usual summation can be replaced by the
Monte Carlo one.

In this article the algorithm based on  Dyson-Schwinger
recursive equations is presented and used in order to  efficiently obtain
cross sections for arbitrary  multi-jet processes.

\section{DESCRIPTION OF THE ALGORITHM}
To illustrate how the algorithm works let us present a simple
example, the gluon self-interaction:
\begin{equation} 
g(p_{1})g(p_{2}) \rightarrow g(p_{3})g(p_{4}).
\end{equation}  
Here $p_{1},p_{2},p_{3},p_{4}$ represent the external momenta involved in the
scattering process, taken to be incoming. 
The subamplitude with an off-shell gluon of momentum $P$ has contributions 
from three- and four-gluon vertices only.
To reduce the computational complexity down to an asymptotic $3^n$,
we replaced the four-gluon vertex
with a three-gluon vertex by introducing an auxiliary field 
represented by the antisymmetric tensor $H^{\mu\nu}$.
The recursion for the gluons now changes only slightly.   However,
we have an additional equation for the auxiliary field. 
\begin{figure}[!ht]
\begin{center}
\epsfig{file=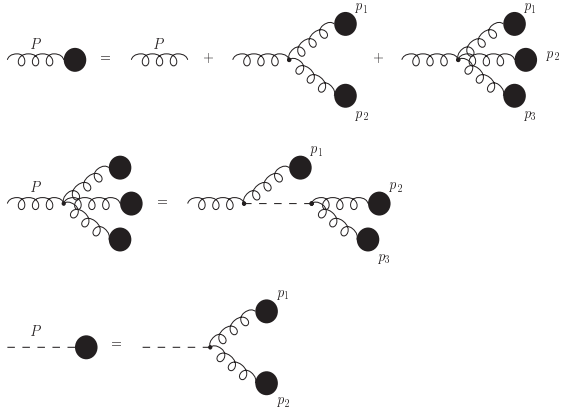,width=75mm,height=55mm}
\end{center}
\vspace{-1 cm}
\caption
{\em Recursion equations for gluon self-interactions. The auxiliary field
is used to replace the four-gluon vertex in order to reduce computational 
complexity of calculations.}
\label{gluons}
\end{figure}
Diagrammatically, the full content of Dyson-Schwinger equations
is presented in  Fig.~\ref{gluons} and the corresponding equations have 
the following form:
\[
A^{\mu}_{AB}(P)=\frac{g_{s}}{2P^{2}}\sum_{P=p_{1}+
p_{2}}V^{\mu}_{\nu\lambda}(P,p_{1},p_{2}) \epsilon(p_{1},p_{2})
\]
\[
\{ A^{\nu}_{AC}(p_{1})A^{\lambda}_{CB}(p_{2})-
A^{\lambda}_{AC}(p_{2})A^{\nu}_{CB}(p_{1})\}
\]
\[
+\frac{ ig_{s}}{2P^{2}}\sum_{P=p_{1}+p_{2}}
X^{\mu}_{\nu\lambda\rho}\{ A^{\nu}_{AC}(p_{1}) H^{\lambda\rho}_{CB}(p_{2})
\]\[
-H^{\lambda\rho}_{AC}(p_{2})A^{\nu}_{CB}(p_{1})\}
\epsilon(p_{1},p_{2})
\]
\[
H^{\mu\nu}_{AB}(P)=\frac{ig_{s}}{4}\sum_{P=p_{1}+p_{2}}X^{\mu\nu}_{\lambda\rho}
~\epsilon(p_{1},p_{2})
\]
\[
\{ A^{\lambda}_{AC}(p_{1})A^{\rho}_{CB}(p_{2})- 
A^{\rho}_{AC}(p_{2})A^{\lambda}_{CB}(p_{1}) \}
\]
where $X^{\mu\nu\lambda\rho}$ is the new auxiliary field-gluon-gluon vertex:
\begin{equation}
X^{\mu\nu\lambda\rho}=g^{\mu\lambda}g^{\nu\rho}-g^{\nu\lambda}g^{\mu\rho}
\end{equation}
and $A,B,C=1,2,3$, and $\epsilon(p_{1},p_{2})$ is the Fermi sign factor.
In order to be more transparent we will write explicitly the three-gluon vertex
part as well as the auxiliary one, suppressing the color indices and
using the light-cone representation:
\[
A^{\mu}(P) \sim
(A(p_{1})\cdot A(p_{2}))(p_{2}-p_{1})^{\mu} +(p_{1}\cdot A(p_{2})+
\]
\[
P \cdot A(p_{2}))A^{\mu}(p_{1})
-(p_{2}\cdot A(p_{1})+P \cdot A(p_{1}) ) 
A^{\mu}(p_{2})
\]
\[
+ A_{\nu}(p_{1})H^{\mu\nu}(p_{2})-A_{\nu}(p_{1})H^{\nu\mu}(p_{2})
\]
\begin{equation}
H^{\mu\nu} (P) \sim A^{\mu}(p_{1})A^{\nu}(p_{2})
-A^{\nu}(p_{1})A^{\mu}(p_{2})
\end{equation}
where all momenta are taken to be incoming.
These equations, the off-shell fields, are the main building blocks
of the gluon self-interaction, which is to be constructed
iteratively. After $n-1$ steps, where
$n$ is the number of particles under consideration,
one can get the total amplitude 
\begin{equation}
{\cal A}(p_{1},p_{2},...,p_{n})= 
{\hat A}_{AB}(P_{i})\cdot A_{AB}(p_{i}),
\end{equation}
where
\begin{equation}
P_{i}=\sum_{j\ne i} p_{j}
\end{equation}
so that $P_{i}+p_{i}=0$. The hat denotes functions  given by
the expression from the previous step except for the propagator
term. This is because 
the outgoing momentum $P_{i}$ must be on shell and the propagator term
is removed by the amputation procedure. The iteration begins with the
initial condition for the external particles. For gluons we have
\begin{equation}
A^{\mu}_{AB}(p_{i})=\varepsilon^{\mu}_{\lambda}(p_{i})\delta_{AC}\delta_{DB},
\end{equation}
where $i$ enumerates the external particles,
$i=1,...,n$, $\varepsilon$ denotes the polarization vector and
$\lambda=\pm 1$.

In order to label and systematically control the
momenta of  
 external particles and their relevant intermediate combination,
we give all momenta in the binary representation, see {\it e.g.}
\cite{Caravaglios:1995cd}. 
 For the process under consideration, 
and the external particles with momenta $p^{\mu}_{i}$, $i=1,2,3,4$ 
we assign to the momentum $P^{\mu}$
a binary vector $\vec{m}=(m_{1},m_{2},m_{3},m_{4})$ with components which 
are either $0$ or $1$ as follows:
\begin{equation}
P^{\mu}=\sum_{i=1}^{4}m_{i}p_{i}^{\mu}.
\end{equation}
The binary vector can now be uniquely represented by the integer
\begin{equation}
m=\sum_{i=1}^{4}2^{i-1}m_{i}.
\end{equation}
In particular, one can write: 
\[
g(1)g(2) \rightarrow g(4)g(8).
\]
All subamplitudes can now be  replaced by the 
corresponding integers 
\[
A^{\mu}(P)\rightarrow A^{\mu}(m),
\]
\[H^{\mu\nu}(P)\rightarrow H^{\mu\nu}(m).
\]
This representation allows us to establish a natural ordering of the momenta
based on the notion of  level, defined simply as
\begin{equation}
l=\sum_{i=1}^{4}m_{i}. 
\end{equation}
All external momenta are at the level $1$, whereas the total amplitude 
corresponds to the unique level $n=4$. With  levels  we can see the
natural iteration path of  
the equations. We start from the subamplitudes at the level $1$ which are 
the external momenta together with the initial conditions, then go to 
the subamplitudes at the level $2$ and so on until we reach level
$n=4$. The Fermi sign factor is also expressed
in the integer number language 
\begin{equation}
\epsilon(P_{1},P_{2})\rightarrow\epsilon(m_{1},m_{2})
\end{equation}
and we use the following formula
\begin{equation}
\epsilon(m_{1},m_{2})=(-1)^{\chi(m_{1},m_{2})}
\end{equation}
with
\begin{equation}
\chi(m_{1},m_{2})=\sum_{i=n}^{2}\hat{m}_{1i}\sum_{j=1}^{i-1}\hat{m}_{2j},
\end{equation}
where the hat means that this particular component is equal to $0$ if the 
corresponding external particle is a boson. This sign factor takes into 
account the  sign change when two identical fermions are interchanged.

Contrary to orginal {\tt HELAC} 
\cite{Kanaki:2000ey,Kanaki:2000ms}, the computational part consists of only one
step, where couplings allowed by the lagrangian defined by 
fusion rules are only
explored, see Ref.~\cite{new} for technical details. Subsequently, the helicity
configurations are set up. There are two possibilities, either exact summation
over all $2^{n_{1}}\times 2^{n_{2}} \times 2^{n_{3}} $ helicity
configurations, where $n_{1}$ is the total number of quarks,
$n_{2}$, the total number of antiquarks and $n_{3}$, the total number 
of gluons; or Monte Carlo summation. 
For example for 
a gluon the second option is achieved by introducing 
the polarization vector  
\begin{equation}
\varepsilon^{\mu}_{\phi}(p)=e^{i\phi}\varepsilon^{\mu}_{+}(p)+
e^{-i\phi}\varepsilon^{\mu}_{-}(p),
\end{equation}
where $\phi \in (0,2\pi)$ is a random number. By integrating over 
$\phi$ we can obtain the sum over helicities 
\[
\frac{1}{2\pi}\int_{0}^{2\pi}d\phi ~\varepsilon^{\mu}_{\phi}(p)
(\varepsilon^{\nu}_{\phi}(p))^{*}=\sum_{\lambda=\pm}
\varepsilon_{\lambda}^{\mu}(p)(\varepsilon_{\lambda}^{\nu}(p))^{*}.
\]
The same idea can be applied 
to the helicity of quarks and antiquarks.

Finally, the color factor is evaluated iteratively. Once again, we
have two options. Either we  
proceed by computing all $3^{n_{q}}\times3^{n_{\bar{q}}}$
color configurations, where the gluon is 
treated as a quark-antiquark pair and $n_{q},n_{\bar{q}}$ is  the number of 
quarks and antiquarks respectively, or 
particular configurations are chosen by the Monte Carlo method (for more
details see Ref.~\cite{new}).

For the process $g(P_{1})g(P_{2}) \rightarrow g(P_{4})g(P_{8})$ 
which we are using through this
 section as an example the momenta involved in the various levels 
of  calculation are the following: 
\begin{flushleft}
{\underline {1st Level:} }
\end{flushleft}
\begin{flushleft}
$P_{1}=(0001)=p_{1}, ~~P_{2}=(0010)=p_{2}$
\end{flushleft}
\begin{flushleft}
$P_{4}=(0100)=p_{3}, ~~P_{8}=(1000)=p_{4}$
\end{flushleft}
\begin{flushleft}
{\underline {2nd Level:} } 
\end{flushleft}
\begin{flushleft}
$P_{6}=(0110)=p_{2}+p_{3} $
\end{flushleft}
\begin{flushleft}
$P_{10}=(1010)=p_{2}+p_{4}$ 
\end{flushleft}
\begin{flushleft}
$P_{12}=(1100)=p_{3}+p_{4}$
\end{flushleft}
\begin{flushleft}
{\underline {3th Level:} } $P_{14}=(1110)=p_{2}+p_{3}+p_{4}$
\end{flushleft}
\begin{flushleft}
{\underline {4th Level:} } $P_{15}=(1111)=p_{1}+p_{2}+p_{3}+p_{4}$.
\end{flushleft}
Note that we have chosen the particle number $1$ as our ending point
and, therefore, we 
have only computed all the relevant momentum combinations where the
momentum $p_{1}$ 
does not appear. This excludes all odd integers between $1$ and $2^{n}-2$.
The momentum $p_{1}$ is combined only once, in the last  $n-$th level. 
Starting from the second level we get:
\[
A^{\mu}(6) \sim V^{\nu\rho\sigma}(6,2,4)
A_{\rho}(2)A_{\sigma}(4) 
\]
\[
H^{\mu\nu}(6) \sim
X^{\mu\nu\rho\sigma}A_{\rho}(2)A_{\sigma}(4)
\]
\[
A^{\mu}(10)\sim V^{\nu\rho\sigma}(10,2,8)
A_{\rho}(2)A_{\sigma}(8) 
\]
\[
H^{\mu\nu}(10)\sim
X^{\mu\nu\rho\sigma}A_{\rho}(2)A_{\sigma}(8)
\]
\[
A^{\mu}(12)\sim V^{\nu\rho\sigma}(12,4,8)
A_{\rho}(4)A_{\sigma}(8) 
\]
\[
H^{\mu\nu}(12)\sim 
X^{\mu\nu\rho\sigma}A_{\rho}(4)A_{\sigma}(8)
\]
At level  $3$:
\[
A^{\mu}(14)\sim V^{\mu\rho\sigma}(14,2,12)A_{\rho}(2)A_{\sigma}(12)
-
\]
\[
X^{\mu\nu\rho\sigma}A_{\nu}(2)H_{\rho\sigma}(12)
+V^{\mu\rho\sigma}(14,4,10)A_{\rho}(4)A_{\sigma}(10)
\]\[
-X^{\mu\nu\rho\sigma}A_{\nu}(4)H_{\rho\sigma}(10)
+V^{\mu\rho\sigma}(14,8,6)A_{\rho}(8)A_{\sigma}(6)
\]\[
-X^{\mu\nu\rho\sigma}A_{\nu}(8)H_{\rho\sigma}(6).
\]
At the last, 4-th level,
the total amplitude is computed by combining this subamplitude 
with the remaining one, which describes the momentum $p_{1}$ and is 
simply given by
\[
{\cal A}(15)\sim A_{\mu}(1)\cdot A^{\mu}(14).
\]

\section{NUMERICAL RESULTS}

\begin{table}[!ht]
\newcommand{\lstrut}{{$\strut\atop\strut$}}
\caption {\em  Results for the total cross section  from $4$ up to $8$ 
gluons, $\sigma_{\textnormal {\tiny T}}$ corresponds to summation over 
all possible color configurations, while 
$\sigma_{\textnormal{{\tiny T}}}^{\textnormal{\tiny MC}}$ corresponds to 
 Monte Carlo summation. }
\vspace{0.2cm}
\label{n-jet}
\begin{center}
\begin{tabular}{||l|l|l||}
\hline \hline & & \\
$\textnormal{Process}$ & $\sigma_{\textnormal {\tiny T}}$ 
$\pm$ $\textnormal{error}$ $\textnormal{(nb)}$ 
& $\sigma_{\textnormal{{\tiny T}}}^{\textnormal{\tiny MC}}$
$\pm$ $\textnormal{error}$ $\textnormal{(nb)}$ \\  & & \\
\hline \hline $gg \rightarrow 2g$ & 4611.55 $\pm$ 38.13  &  
4627.18 $\pm$ 33.28 \\ 
\hline  $gg \rightarrow 3g$ & 152.444 $\pm$ 2.490 & 
152.137 $\pm$ 2.822  \\ 
\hline  $gg \rightarrow 4g$ & 12.9072 $\pm$ 0.4070  & 12.6137 $\pm$ 0.4619\\
\hline $gg \rightarrow 5g$ & 1.04254 $\pm$ 0.05300 & 1.04446 $\pm$ 0.10390  \\
\hline $gg \rightarrow 6g$ & 0.07577 $\pm$ 0.00597 &   0.07261 $\pm$ 0.00516 \\

\hline\hline
\end{tabular}
\end{center}
\end{table}
\begin{figure}[!ht]
\begin{center}
\epsfig{file=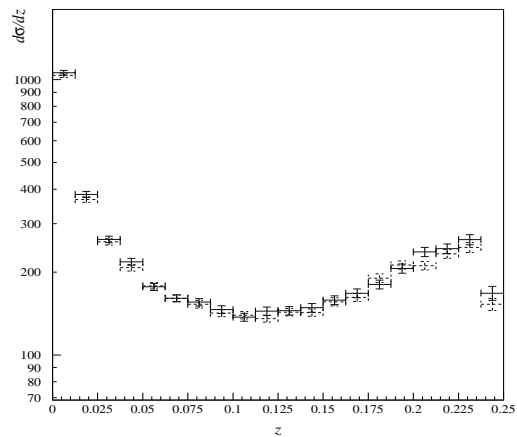,width=75mm,height=65mm}
\epsfig{file=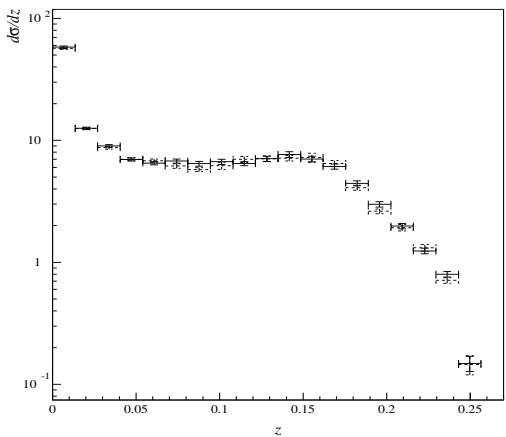,width=75mm,height=65mm}
\end{center}
\vspace{-1cm}
\caption{\em Distribution in $z=|M_{I}|^2/\sum_{i}^{all} |M_{i}|^{2}$,
where $|M_{I}|^2$ is the square matrix element for one particular color 
configuration normalized to the sum of all possible is plotted. The
upper plot corresponds to the  
 $gg \rightarrow gg$ process, the lower one to the $gg \rightarrow ggg$. 
Solid line crosses denote summation over all color configurations
whereas dashed, the Monte Carlo summation. }
\label{4g_5g_flows}
\end{figure}

As an example the algorithm has been used to compute total cross sections 
for multiple  jets production. 
The CMS energy was chosen $\sqrt{s}=14$ $\textnormal{TeV}$ and
the following cuts were applied
\begin{equation}
p_{T_{i}} > 60 ~\textnormal{GeV}, ~~~~ |\eta_{i}|<2.5, ~~~~ \Delta R > 1.0
\end{equation}
where $p_{T}=\sqrt{p_{x}^{2}+p_{y}^{2}}$
is the transverse momentum of a jet, $\eta=-\ln \tan(\theta /2)$  is the
 pseudorapidity. Many methods can be used to define 
what is meant by a jet of hadron. One commonly used is the 'cone' 
description of a jet which is the transverse energy, $E_{T}$, 
concentration in a cone of radius  
\[
\Delta R=\sqrt{\Delta\Phi_{i}^{2}+\Delta\eta_{ij}^{2}}
\]
with
\[ 
\Delta \Phi_{ij}=\arccos \biggl( \frac{p_{x_{i}}p_{x_{j}}+p_{y_{i}}p_{y_{j}}}
{p_{T_{i}} p_{T_{j}} } \biggr).
\]
All results are obtained with a fixed strong coupling constant calculated 
at the $M_{Z}$ scale.
There are several parameterizations for the parton structure functions, we 
used  {\tt CTEQ6 PDF}'s parametrization \cite{Pumplin:2002vw,Stump:2003yu}. 
For the phase space generation we used either {\tt PHEGAS}
\cite{Papadopoulos:2000tt} 
or a flat phase-space generator {\tt RAMBO} \cite{Kleiss:1985gy}.

The results for the total cross section  from $4$ up to $8$ 
gluons are listed above in Table~\ref{n-jet}. We give the 'exact'
result with summation over all possible color configurations  
($10^6$ generated events), 
$\sigma_{\textnormal{{\tiny T}}}$   as well as the result obtained
with  Monte Carlo
summation over color ($4\times 10^6$ generated events)  
$\sigma_{\textnormal{{\tiny T}}}^{\textnormal{\tiny MC}}$. In both
cases a Monte Carlo over helicity is applied. 

The next result we present is an example of summation over all color 
configurations in comparison to Monte Carlo  summation. In
Fig.~\ref{4g_5g_flows}, the X-axis variable is
\[
z=|M_{I}|^2/\sum_{i}^{all} |M_{i}|^{2}
\]
where $|M_{I}|^2$ is the square matrix element for one particular color 
configuration normalized to the sum of all possible is plotted. 
The agreement is easy visible.

\section{SUMMARY}
An efficient tool for  automatic computation of helicity 
amplitudes and cross sections for multi-particle final states in 
 QCD has been presented.   

\section*{Acknowledgments}
Work supported by the Polish State Committee for Scientific 
Research Grants number 1 P03B 009 27 for years 2004-2005
(M.W.).  In addition, 
M.W. acknowledges the Maria Curie Fellowship granted by the 
the European Community in the framework of the Human Potential Programme 
under contract HPMD-CT-2001-00105 
({\it ``Multi-particle production and higher order correction''}).

\providecommand{\href}[2]{#2}

\end{document}